\documentclass[
aps,%
10pt,%
final,%
notitlepage,%
oneside,%
twocolumn,%
nobibnotes,%
nofootinbib,% 
superscriptaddress,%
noshowpacs,%
centertags]%
{revtex4}

\usepackage{aas_macros}
\usepackage{framed}
\usepackage{graphics}
\usepackage{graphicx}
\usepackage{upgreek}

\begin{document}
\selectlanguage{english}

\title{Ya. B. Zeldovich and foundation of\\
the accretion theory}

\author{\firstname{Nikolay~I.}~\surname{Shakura}}
\email{nikolai.shakura@gmail.com}
\affiliation{%
Sternberg Astronomical Institute, Moscow M.V. Lomonosov State University, 13 Universitetskij pr.,  Moscow 119992, Russia;\\
Kazan Federal University, Kremlevskaya 18, 420008 Kazan, Russia
}%

%\date{\today}
%\today

\begin{abstract}
This short review is dedicated to academician Yakov Borisovich
Zeldovich, the science of his epoch and the creation of modern accretion theory.
\end{abstract}

\maketitle

\section{Introduction}

Yakov Borisovich Zeldovich was born on March 8, 1914, in Minsk.
His father Boris Naumovich Zeldovich was a lawyer, a member of  a College of Lawyers.
His mother Anna  Petrovna Zeldovich (Kiveliovich) was a translator, a member of the Union of Writers.

\begin{figure}
\centering
\includegraphics[angle=0,width=0.45\textwidth]{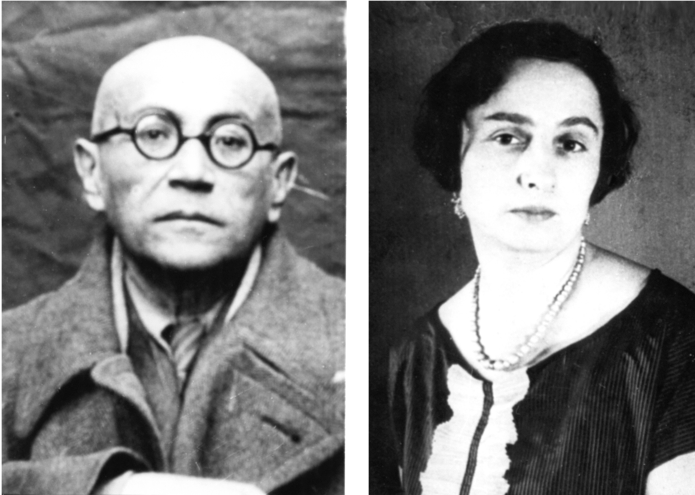}
\caption{
Parents of Ya.~B.~Zeldovich.
}
\label{fig.parents}
\end{figure}

My first acquaintance with Ya.~B.~Zeldovich began with buying of his
book ``Further mathematics for beginners''.
In the middle school my math teacher Alfred Viktorovich Baranovskiy explained
us a method of finding extremum of parabola using Vieta's formulas.
And he told us by the way that using of the methods of further mathematics
allow to make this process easier and more elegant. He did not tell any
details and I was intrigued. And suddenly in a book store of Bobruisk city
(no far from my town) I saw that  Ya.~B.~Zeldovich's math book.
I bought it hoping to find out the method of finding extrema of parabola.
But few days later I went to pass entrance exams to Faculty of Physics
of the Moscow University.

\begin{figure}
\centering
\includegraphics[angle=0,width=0.45\textwidth]{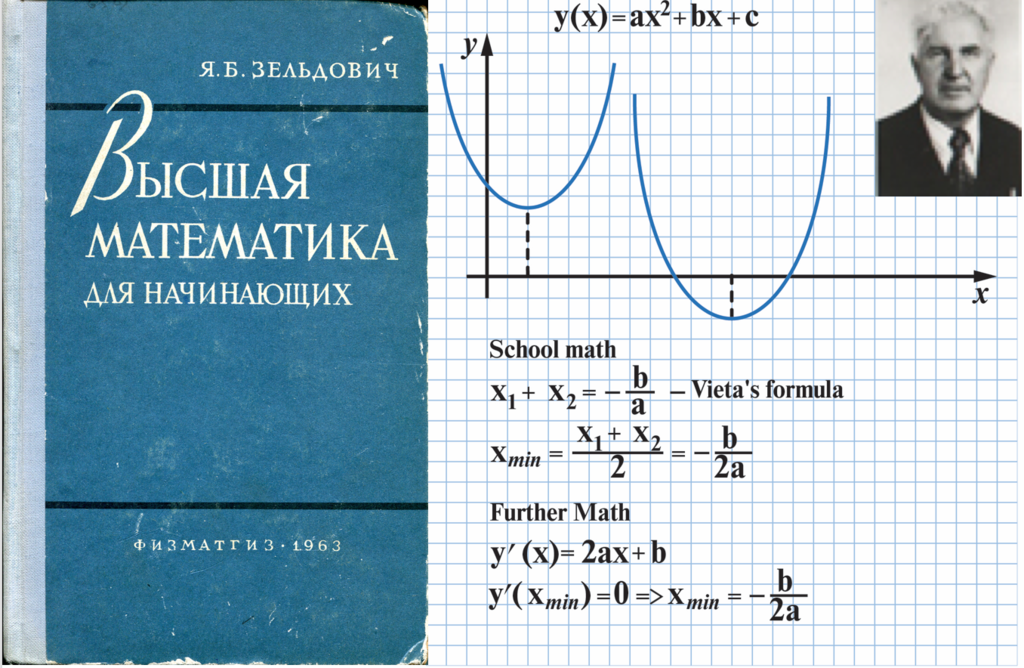}
\caption{
Left panel is Zeldovich's handbook ``Further mathematics for beginners''. Right panel is parabola extrema derivation using two methods: school method (that my teacher A.~V.~Baranovskiy taught me) and further math method from Zeldovich's book. Right upper corner is the photo of Alfred Viktorovich Baranovskiy.
}
\label{fig.further_math}
\end{figure}

In 1963 I entered Astronomical Division of the Lomonosov Moscow State University (MSU).
My decision to study astronomy was inspired by another book that
somehow found its way to Belarus of my childhood:
``Essays About The Universe'' written by Boris Aleksandrovich Vorontsov-Velyaminov,
a professor of astronomy at MSU.
Later I attended his lectures as MSU student and was
examined by him --- something I had never even dreamed about it two or three years ago
when I studied his textbook ``Astronomy'' in school.

\begin{figure}
\centering
\includegraphics[angle=0,width=0.25\textwidth]{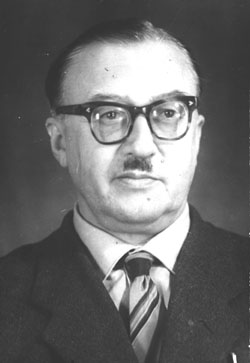}
\caption{
Professor B.~A.~Vorontsov-Velyaminov.
}
\end{figure}

In my first three years at the MSU, Ya.~B.~Zeldovich was nowhere in my surroundings,
nor did I ever remember the book I had bought in Bobruisk.
This book was not included in the list of recommended texts --- by no means
because it was bad, but because the academician addressed it to
beginning engineers and technicians as a self-education book on mathematics.
Academician Ya.~B.~Zeldovich is known as an author of more
than ten high-quality books on a wide range of topics in the sciences.

\begin{figure}
\centering
\includegraphics[angle=0,width=0.45\textwidth]{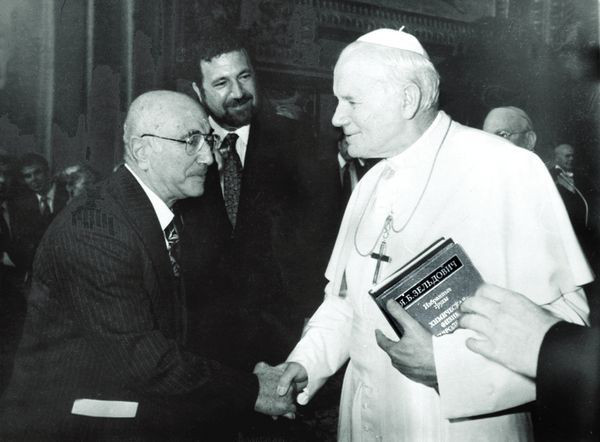}
\caption{
Ya.~B.~Zeldovich presents two values of his selected works to Pope John Paul~II.
At the center of the photo -- professor Remo Ruffini, ICRANet director.
}
\end{figure}

For the first time I met Jakov Borisovich when the Dean's office of Faculty of Physics
organized a meeting of students with the editorial board of the famous Soviet
scientific journal Physics Uspeckhi.
It was the third year of my studying in MSU.
The Large Physics Auditorium being the venue.
I was, of course, very impressed with the brilliance of Eduard Vladimirovich
Shpolsky, the Editor-in-Chief.
Yakov Borisovich Zeldovich was silent all the time, sitting with his densely
hairy hands crossed under his chin and being deep in his own minds.
Later on, already as his co-worker, I found out that he had never been a
full-time undergraduate student at any higher education institution.
Having graduated from a ten-year school in Leningrad in 1930,
he started to work as a laboratory technician at the Institute of Mechanical
Processing of Minerals for some time and then at the Institute of Chemical
Physics (ICP) where, at the age of twenty (!), he began his postgraduate
research under supervising of Nikolay Nikolayevich Semyonov and where it took
him a fantastically short time to get to the very top of the academic career
ladder.

\begin{figure}
\centering
\includegraphics[angle=0,width=0.45\textwidth]{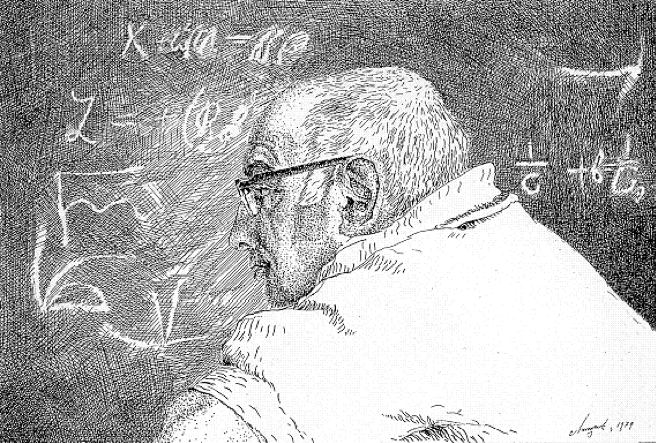}
\caption{
Graphics portrait of Ya.~B.~Zeldovich by V.~M.~Lipunov, early 1980s.
}
\end{figure}

\begin{figure}
\centering
\includegraphics[angle=0,width=0.25\textwidth]{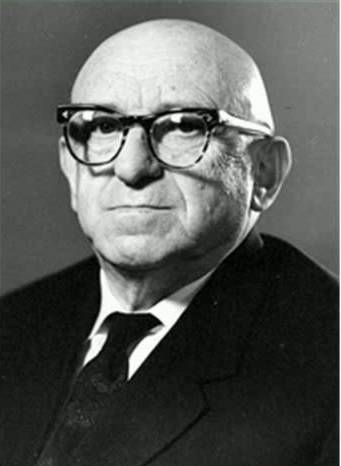}
\caption{
Professor E.~V.~Shpolsky
}
\end{figure}

\begin{figure}
\centering
\includegraphics[angle=0,width=0.25\textwidth]{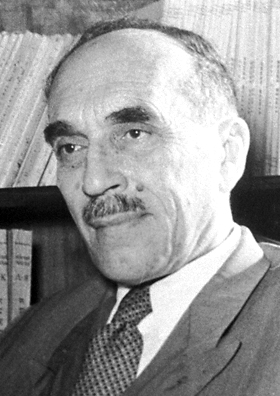}
\caption{
Academician N.~N.~Semyonov, Nobel prizewinner.
}
\end{figure}

In late 1930-s Soviet Journal of Experimental and Theoretical Physics
published several papers about uranium fission by Ya.~B.~Zeldovich and
Yuliy Borisovich
Khariton~\cite{1939ZhETF...9.1425Z,1940ZhETF..10...29Z,1940ZhETF..10..477Z}.
In the same time two papers about a division of
uranium nuclei by free neutrons were published: experimental paper by O.~Hahn and
E.~Strassmann~\cite{1939NW.....27...11H},
and theoretical paper by L.~Meitner and O.~R.~Frisch~\cite{1939Natur.143..239M}.
These papers marked the begging of the new era of civilisation development: the creation of powerful nuclear weapons.
For the long time (1948--1965) Ya.~B.~Zeldovich was one of the leading member of
the scientific team of developing Soviet atomic project.
When he was asked about that time he drew a big black square --- a ``Black square''
by Zeldovich.

\begin{figure}
\centering
\includegraphics[angle=0,width=0.45\textwidth]{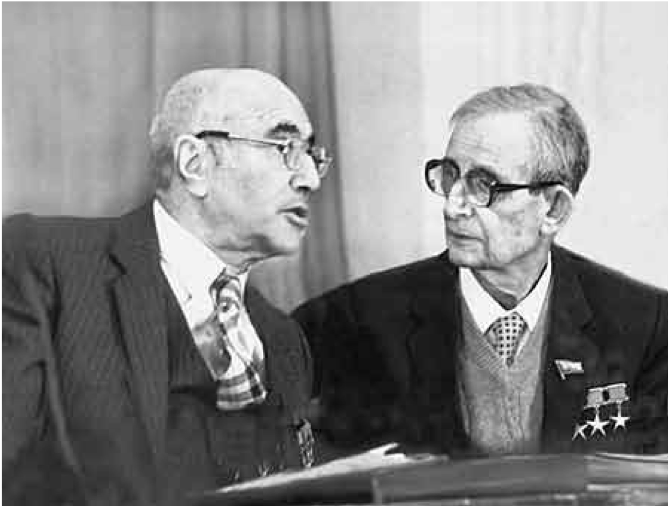}
\caption{
Academicians Y.~B.~Zeldovich and Y.~B.~Khariton, both of them
were awarded three times as Heroes of Socialist Labour.
}
\end{figure}

\begin{figure}
\centering
\includegraphics[angle=0,width=0.225\textwidth]{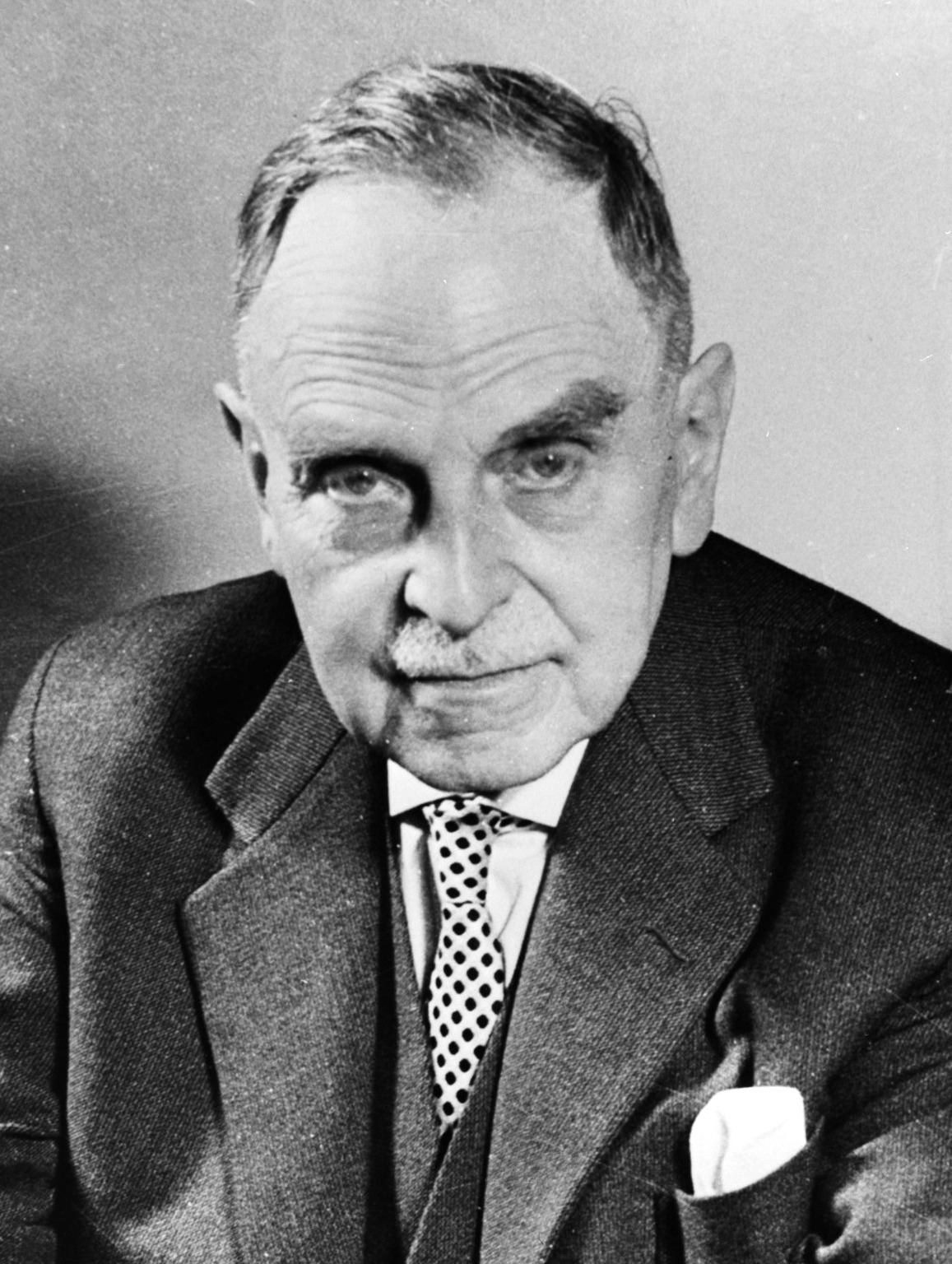}
\includegraphics[angle=0,width=0.225\textwidth]{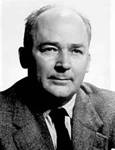}
\includegraphics[angle=0,width=0.225\textwidth]{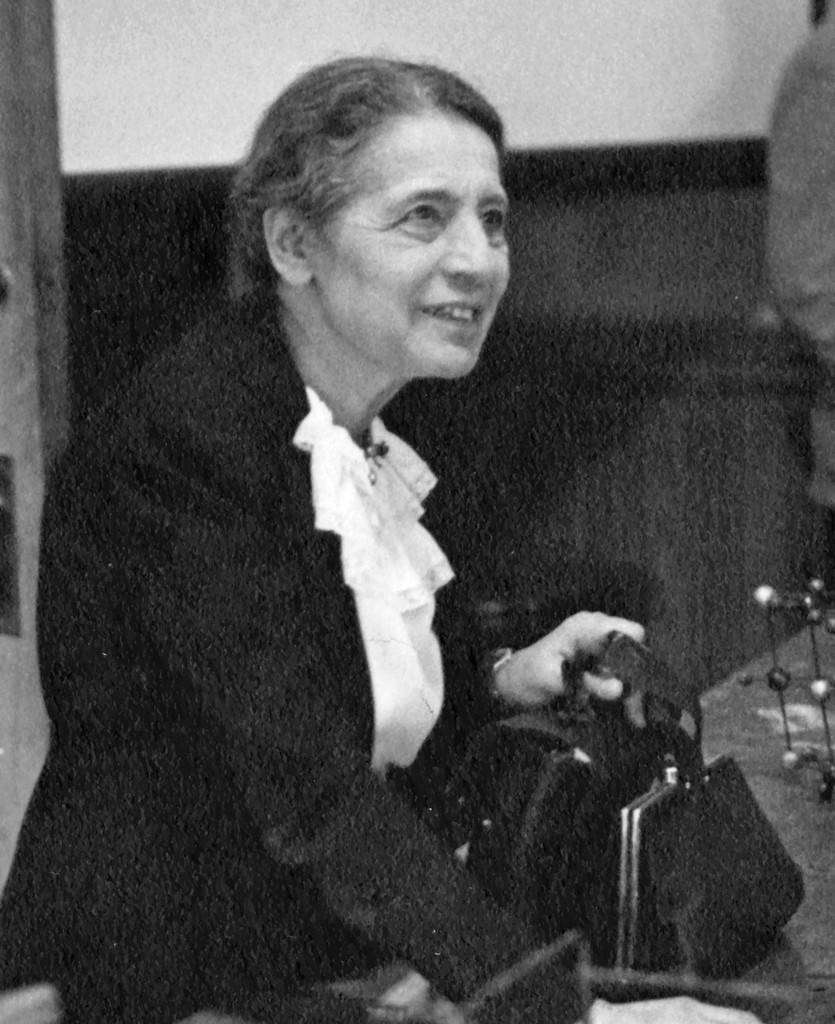}
\includegraphics[angle=0,width=0.225\textwidth]{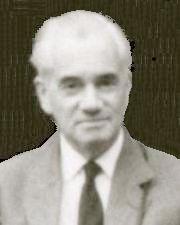}
\caption{
Nobel prizewinner Otto Hahn and his colleagues: professors Fritz Strassmann, Lisa
Meitner and Otto Frisch.
}
\end{figure}

\begin{figure}
\centering
\includegraphics[angle=0,width=0.3\textwidth]{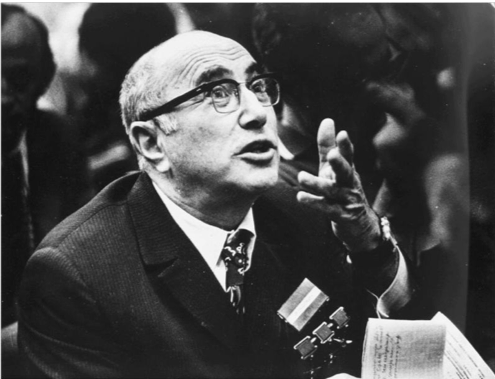}
\includegraphics[angle=0,width=0.3\textwidth]{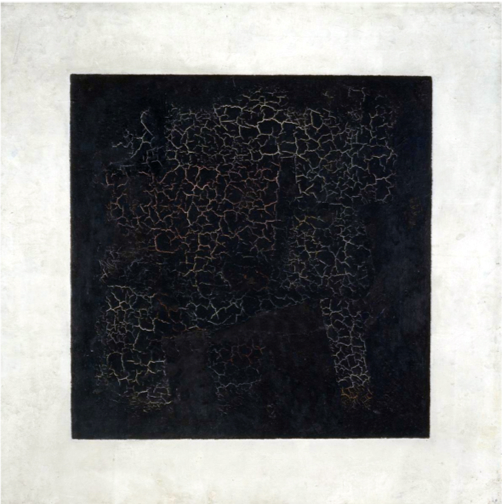}
\caption{
Ya.~B.~Zeldovich with three golden medals of the Hero of Socialist Labour,
that he got for Soviet atomic project. Malevich's Black Square as a symbolic
answer for questions about Zeldovich's participation in the atomic project.
}
\end{figure}

In the September of 1966, we, the fourth-year astronomy undergraduates
of the Faculty of Physics, found ourselves enrolled in a new
lecture course, ``The structure and evolution of stars'', to be taught by
Ya.~B.~Zeldovich.
Namely, at these lectures my first personal contact with the academician 
occurred.
The lectures were held on Fridays, and on Thursdays Ya.~B. (some of other
academicians called Yakov Borisovich so) led a joint
astrophysical seminar (JAS) at the Sternberg Astronomical Institute (SAI MSU),
both for full-fledged scientists and for young higher education graduates.
Undergraduates did not have this seminar in their schedule and could only
drop in on when possible.
When Ya.~B. finished his first lecture, he asked if there were in the audience
students wishing to receive topics for their course theses and could they please
stay on for a while.
I was among those few who did.
When my turn came, he asked me whether I had been at the JAS session
the previous day and whether I had heard the talk on the
(then mysterious) sources of cosmic X-ray radiation, and when I twice said yes,
he made a suggestion.
He offered me to calculate the structure and spectrum of the radiation of a
strong shock wave that arises near the surface of a neutron star due to gas
falling onto the star.

\section{Beginnings of X-ray astronomy}

In 1962, a group of American scientists led by Prof. Riccardo Giacconi discovered
first X-ray sources.
Before that, astronomers had known the only one X-ray source of
extraterrestrial origin, namely, the solar corona.
The coronal gas heated to million degrees due to some unknown mechanisms produces
X-ray emission.
The luminosity of the solar corona in X-rays is approximately one millionth
of the optical luminosity of the Sun ($4\times 10^{33}$ erg/s).
So it was natural to assume that other stars are surrounded by hot coronas too.
However, simple calculations showed that detectors available at that time
could not reveal coronas even around the nearest stars on distances of
several parsecs.

Nevertheless, astronomers tried to detect X-ray radiation from the Moon!
The Moon has no atmosphere, but likely some radiation can be produced as a
luminescence of the Moon's surface being attacked by the solar X-rays and
charged particles.
Thus, precisely at midnight of June 18, 1962, when the full Moon was shining,
the ``Aerobee'' rocket was launched.
It reached a height of 225 km, its flight continued for 350~s and was quite
successful: two of the three Geiger counters, with large surface and good
sensitivity in the range  1.5$-$6 keV, were operating during the flight time.
In this energy range, the Earth's atmosphere is totally opaque.
Suddenly, instead of  X-ray radiation from the Moon, a bright and before unknown
source was discovered, which was far beyond the Solar system
in the direction of  the Scorpion constellation.
It was named Sco X-1.

In the following years, new rocket flights brought more and more discoveries
of new X-ray sources.
Gradually, a sky map covered with X-ray sources of different nature was created.
First  sources got their names according to  directions of their location
(Cyg X-1, Cyg X-2, Her X-1, Cen X-3, and so on).
Later it was revealed that their X-ray luminosities were thousands
or even tens of thousand times stronger than the solar luminosity in the
optical range.
The epoch of X-ray astronomy, the epoch of stunning discoveries
in the Universe began.

According to own Ya.~B.~Zeldovich's simple estimations, the shock wave
originating when the gas surrounding a neutron star falls onto its surface
should produce radiation primarily in the X-ray range.
My goal was to carry out the full calculation and investigate the process in
detail.
The main difficulty was connected with the following circumstance: the
free-path length of a falling particle  near the neutron star's surface is
much greater (tens of times) than the
characteristic timescale of interaction between the radiation and matter.
Usually it is not necessary to calculate the structure of the
shock wave: it is sufficient to specify a change in density, pressure,
temperature, and other physical parameters depending on the velocity and the
adiabatic index of the falling gas.
In my problem, the  density, temperature and other parameters depended on
energy release in the braking zone.
Moreover, plasma oscillations may arise in this zone.
To describe them, a consideration of kinetic plasma equations is required
instead of ordinary hydrodynamics.
Finally, it was shown that shock wave emission spectra
from accreting neutron stars could explain observational data obtained with the
launched equipment.

This investigation was outlined in the article written by Ya.~B. and me, the article
was submitted to ``Soviet Astronomy'' in 1968 and published
in 1969~\cite{1969AZh....46..225Z}.
Meanwhile, in 1967 first radio pulsars were discovered which turned to be a strongly
magnetized rotating neutron stars~\cite{1968Natur.217..709H}.
This discovery starts new epoch of radio astronomy: the epoch of study of various radio
pulsars.
But here I will discuss accreting X-ray pulsars!

\begin{figure}
\centering
\includegraphics[angle=0,width=0.225\textwidth]{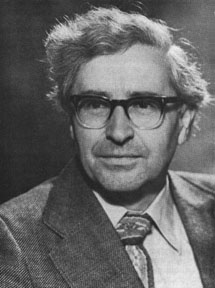}
\includegraphics[angle=0,width=0.225\textwidth]{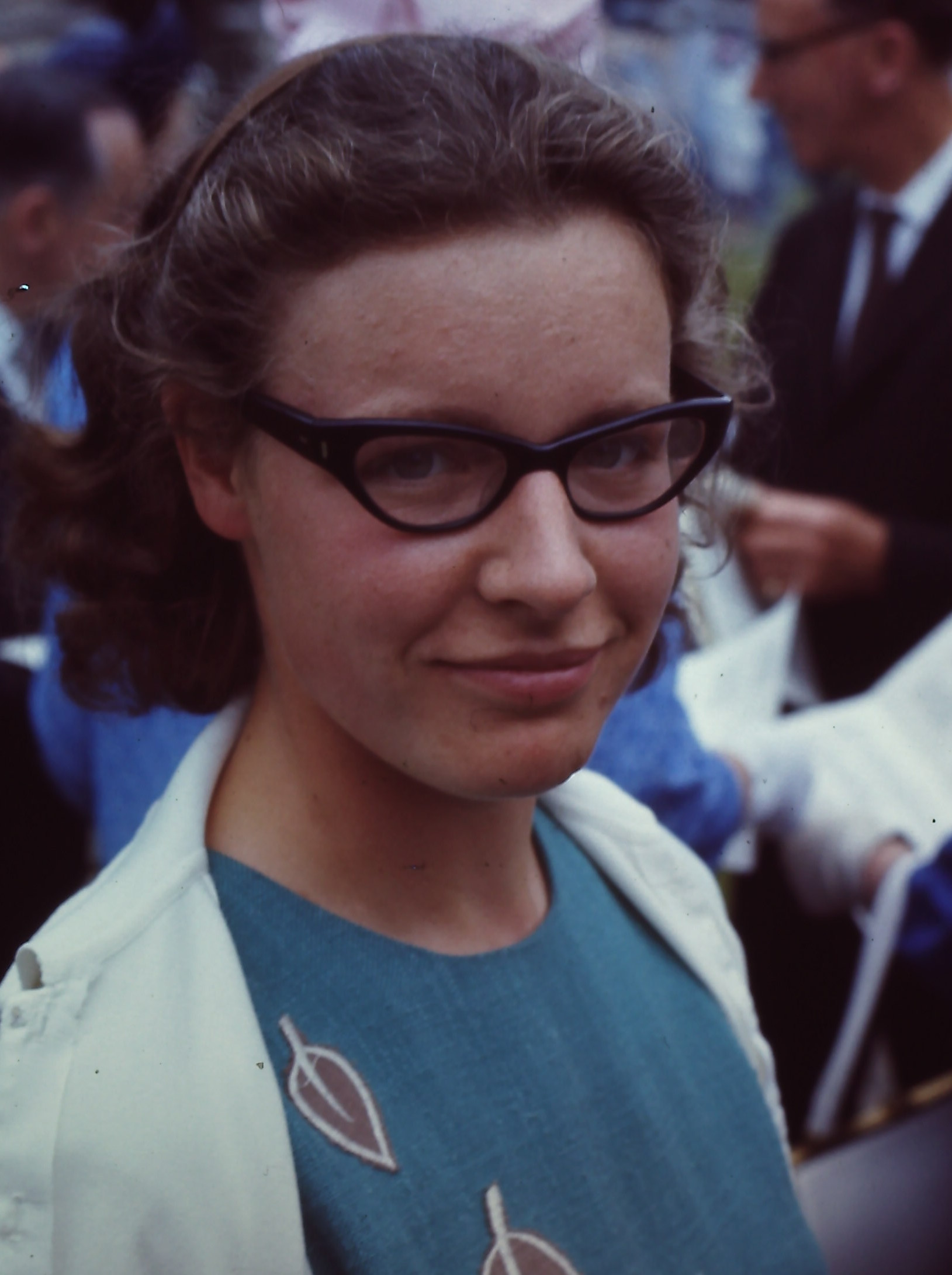}
\caption{
Nobel prizewinner Antony Hewish and his doctoral student Jocelyn Bell.
}
\end{figure}

As an undergraduate, I attended lectures on general astrophysics by Dmitrii
Yakovlevich Martynov, then a Director of SAI, who paid special attention to
close binary stellar systems with the matter flowing from one star to another.
Due to the relative orbital motion of the two stars,
this process results in an origin of a disk-shape envelope around one of these stars.
It seemed natural to consider a binary stellar system where one of the components is
a neutron star or even a black hole.

\begin{figure}
\centering
\includegraphics[angle=0,width=0.25\textwidth]{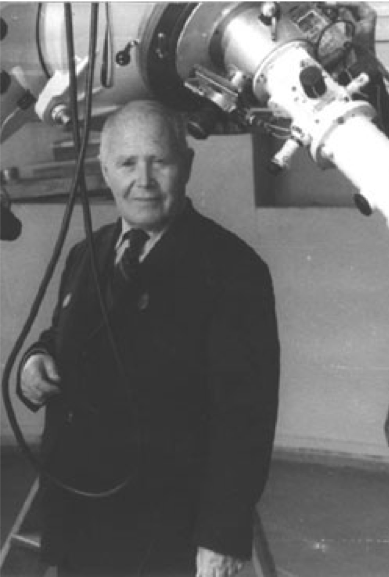}
\caption{
Professor D.~Ya.~Martynov.
}
\end{figure}

\begin{figure}
\centering
\includegraphics[angle=0,width=0.45\textwidth]{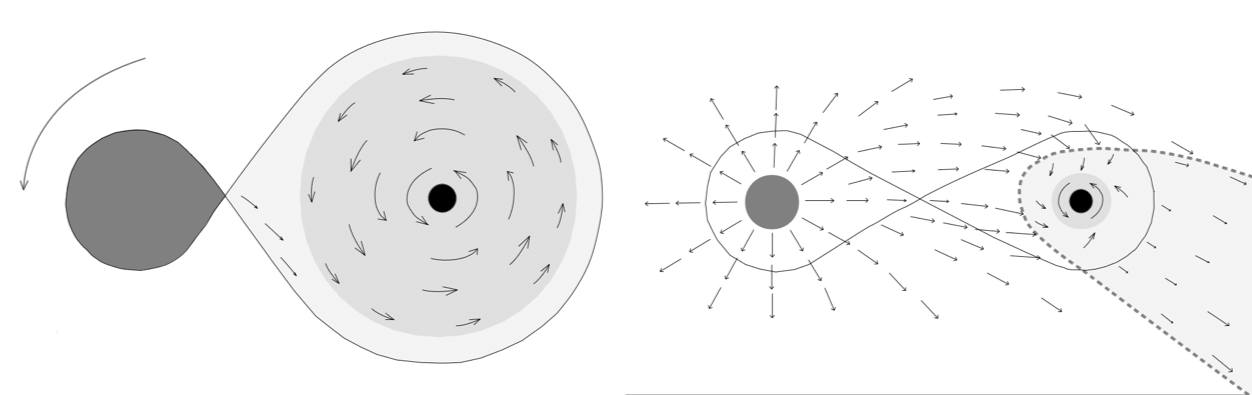}
\caption{
Two types of accretion disk formation in close binary systems with black holes or
neutron stars.
}
\label{fig.disk}
\end{figure}

Thus, imagine a binary stellar system consisting of a normal star and a black hole.
The size of the ordinary star in this system is limited by that of the so-called
Roche lobe.
The normal star can increase in size as stellar evolution proceeds,
and after the Roche lobe is filled, the matter starts flowing from its surface
to the region of gravitational attraction of the black hole (left panel of
Fig.~\ref{fig.disk}).
Because of the relative orbital motion of the binary components, the matter does not
fall directly onto the black hole but forms a differentially rotating disk-like
envelope around it.
Due to layer-to-layer friction, the matter that accumulates in the disk strongly
heats up and starts glowing.
In its rapid revolution, the matter in the disk slowly moves radially
toward (or accretes upon) the black hole, loosing its angular momentum in the process.
The glowing of the disk is caused by the matter accretion releasing its
gravitational energy.
Indeed, the inner parts of the disk closest to the black hole become hot
enough to emit in the X-ray range.

In a more complex accretion disk formation scenario, the optical
companion does not fill its Roche lobe and flows outward in all directions
via stellar wind.
In this case, a head-on shock is naturally expected to form in the region
where the stellar wind stream is under the gravitational influence of the black hole.
After the passage of the shock wave, the matter in the gravitational capture region
of the black hole starts falling onto it -- but not strictly radially!
Due to its rotational motion, the falling matter has a specific angular momentum,
which is somewhat greater than the specific orbital angular momentum of the hole.
In the case of the falling with conserved angular momentum the matter acquires
the orbital motion at some distance from the black hole (right panel of Fig.~\ref{fig.disk}).
And then, well, disk type accretion again!

Replacing the black hole by a strongly magnetized neutron star in the binary system
has a consequence that the stellar magnetic field starts destroying the
accretion disk at a distance of about a hundred star radii.
The accreting matter then starts rapidly falling along the magnetic force lines,
encountering the surface of the neutron star in the vicinity
of the magnetic poles.
Because magnetic and geographic poles are usually far apart,
the rotation of the neutron star causes it to be observed as an accretion pulsar.

The presence of accreting black holes and neuron stars in binary stellar systems
was first detected in the early 1970s by an experiment aboard the US purpose-built
satellite {\sc Uhuru}.
The satellite was launched into a circular orbit about 500 km
high from the Italian marine platform San Marco off the coast of Kenya.
The satellite got name {\sc Uhuru} (which is Swahili for freedom) due to the fact
that the launch date, 12 December 1970, was Kenya's Independence Day.
Recognition is not always quick to come in science, and it was only in 2002
that the mission project leader Riccardo Giacconi, a US astrophysicist of
Italian origin, was awarded the Nobel Prize in Physics for this pioneering work.
A lot of discovered by {\sc Uhuru} X-ray sources are turned to be accreting black holes
and neutron stars.

\begin{figure}
\centering
\includegraphics[angle=0,width=0.15\textwidth]{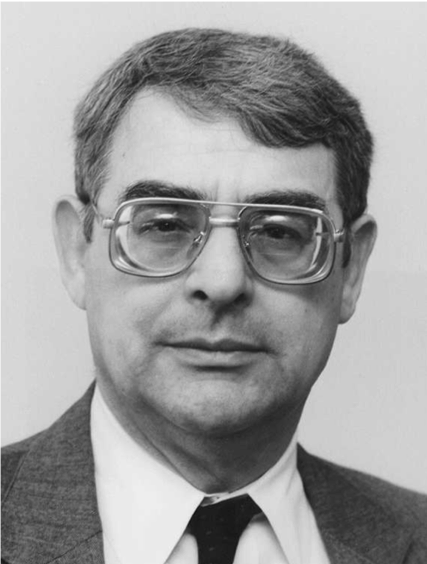}
\includegraphics[angle=0,width=0.30\textwidth]{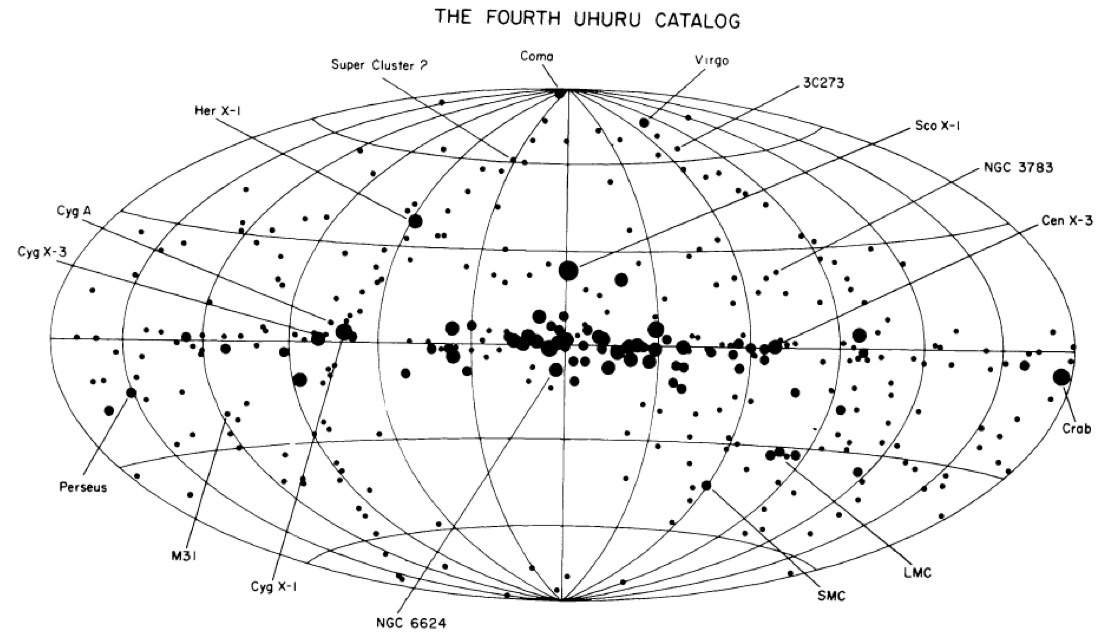}
\caption{
Nobelprize winner Riccardo Giacconi and the map of X-ray sky obtained by {\sc Uhuru}
satellite~\cite{1976ASIC...28..229G}.
}
\end{figure}

Virtually simultaneously with the discovery of accreting black holes and neutrino stars
in binary stellar systems, the foundation of the theory of disk accretion on
gravitating centers was laid by me under the guidance of Ya.~B.
The paper was submitted to ``Soviet Astronomy'' in June 1971 and published
in November 1972~\cite{1972AZh....49..921S}.
Remarkable that the first publication with the results of {\sc Uhuru} was received by
editors of Astrophhysical Journal
in 17th May 1971 and was published on July 1971~\cite{1971ApJ...167L..67G}.

The bulk of the work was done together with Rashid Sunyaev.
In another of our joint efforts, the so-called standard model of disk accretion
was developed and worked out in detail, which was presented~\cite{1973IAUS...55..155S}
at the 55th Symposium of the International Astronomical Union held in
May 1972 in Madrid.
The symposium covered not only data from {\sc Uhuru} but also the first theoretical results
on modeling the compact X-ray sources detected in binary stellar systems,
i.e., accreting black holes and neutron stars.
Our joint work was presented by Jim Pringle (UK) -- Rashid and I were then
``nevyezdnye'' (not free to go abroad) -- and was in fact an introduction to a large
scale paper~\cite{1973A&A....24..337S} which was later published in 1973 in the 
high-profile European journal Astronomy and Astrophysics, and on the basis of which
Igor Dmitrievich Novikov and Kip Thorne were able to calculate accurately the relativistic
corrections to radiation due to general relativity effects near
black holes~\cite{1973blho.conf..343N}.

\begin{figure}
\centering
\includegraphics[angle=0,width=0.45\textwidth]{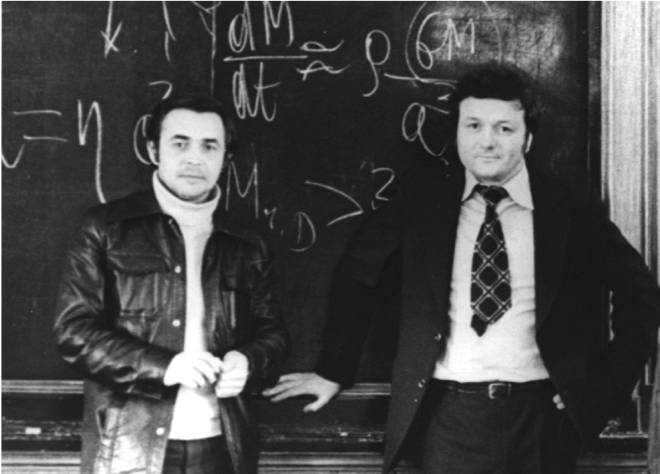}
\includegraphics[angle=0,width=0.45\textwidth]{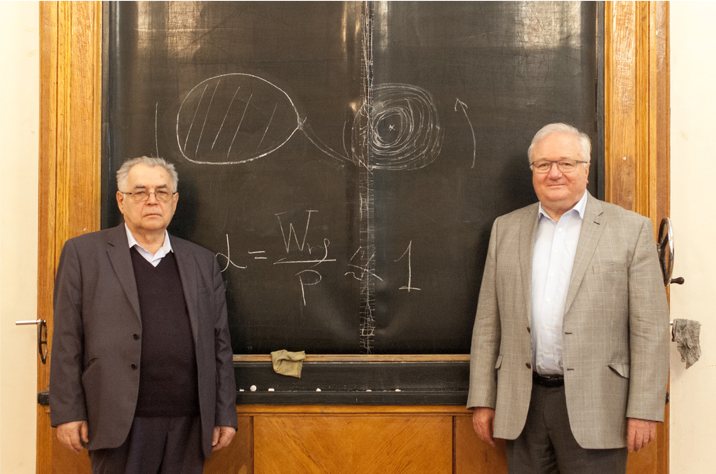}
\caption{
Me and academician Rashid Alievich Sunyaev near the same blackboard in SAI,
70s and 2017.
}
\end{figure}

\begin{figure}
\centering
\includegraphics[angle=0,width=0.45\textwidth]{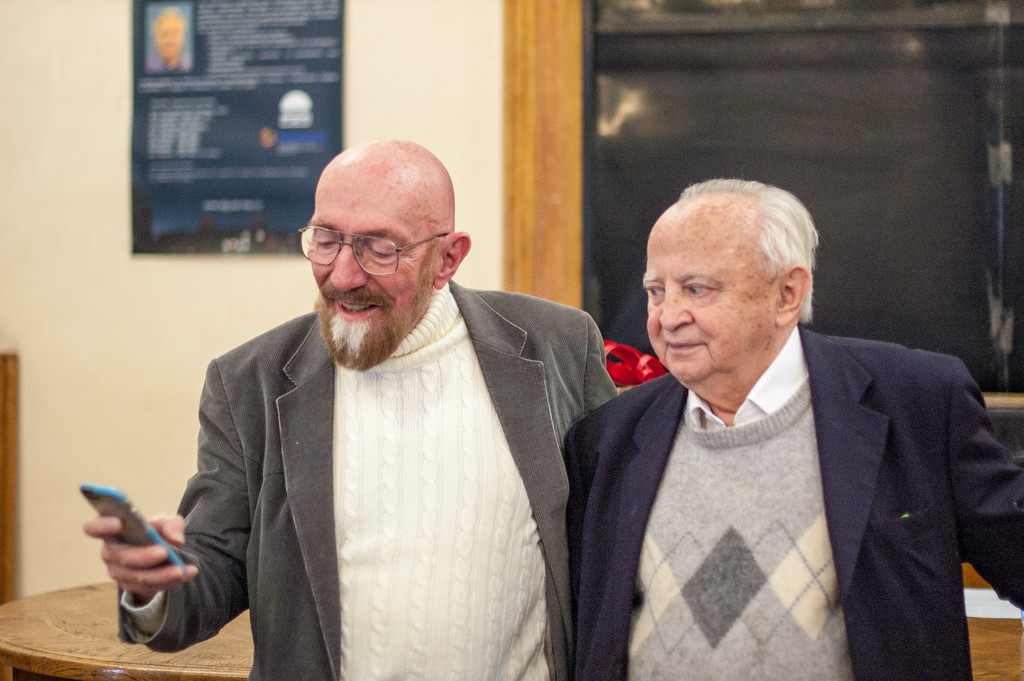}
\caption{
Nobelprize winner K.~Thorne and corresponding member of the Russian Academy of Sciences
I.~D.~Novikov on the Prof. Leonid Petrovich Grishchuk memorial conference, November 2016.
}
\end{figure}

The results {\sc Uhuru} produced during its three years of operation were spectacular.
Not only a large number (339) of newly discovered X-ray sources were catalogued,
but {\sc Uhuru} also provided guidance for other space observatories.
Currently, space X-ray sources of various natures (not necessarily accreting
relativistic stars in binary stellar systems!) number in the hundreds of thousands.

Rashid's and my pioneering paper~\cite{1973A&A....24..337S} gained wide popularity;
its number of citations as of July 2018,
i.e., forty five years on, exceeding 8500.

\begin{figure}
\centering
\includegraphics[angle=0,width=0.45\textwidth]{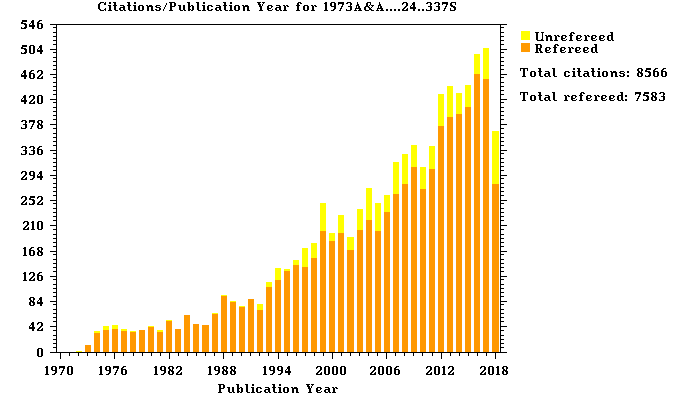}
\caption{
Citation score of \cite{1973A&A....24..337S}, NASA ADS.
}
\end{figure}

The study of accretion disks has led to the discovery in the cores of active
galaxies and in quasars of supermassive black holes with masses ranging from
a few dozen to a few hundred millions solar masses.
The first theoretical results on the glowing of accretion disks around supermassive
black holes were published by David Lynden-Bell (UK) in 1969
in Nature~\cite{1969Natur.223..690L}.

\begin{figure}
\centering
\includegraphics[angle=0,width=0.25\textwidth]{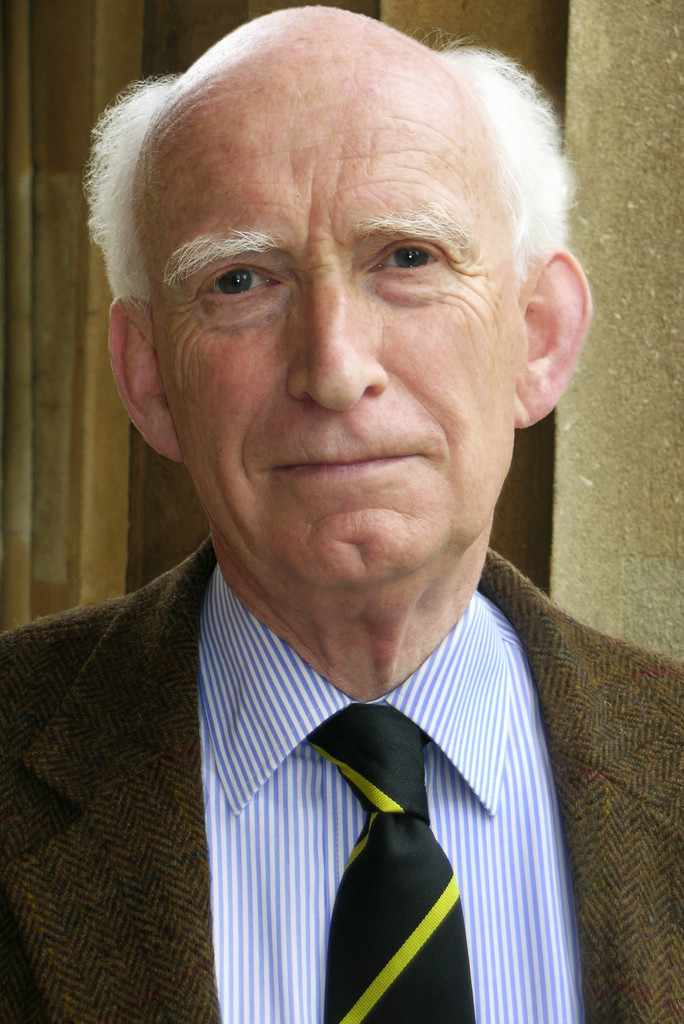}
\caption{
Professor Donald Lynden-Bell.
}
\end{figure}

In the last 45 years a great advance has been made in observations,
as well as in the theory of accretion disks.
Now accretion disks is a specific extensive branch of astrophysics,
important to the same degree as the structure and evolution of stars,
interstellar medium, etc.
The basis for the modern theory of accretion disks is: a) cosmic (magneto)hydrodynamics,
b) radiative transfer, c) general relativity effects in the vicinity of compact stars.

But what about Moon?
``Did R.~Giacconi and his colleagues miss their target?'', you can ask me.
Lunar X-ray light was discovered later by satellite {\sc ROSAT}, that contains more perfect
X-ray telescope with grazing incidence mirrors.

\begin{figure}
\centering
\includegraphics[angle=0,width=0.25\textwidth]{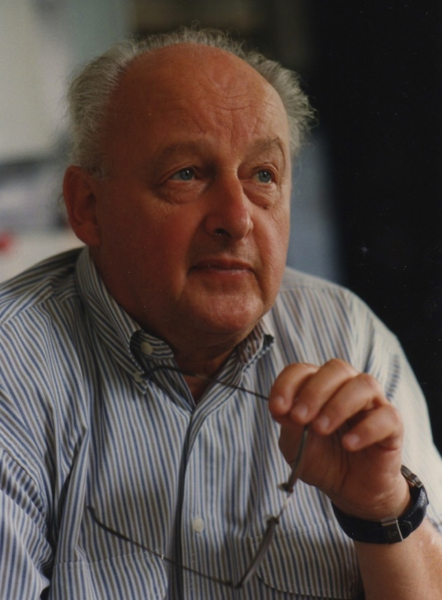}
\caption{
Professor Joachim Trumper, the PI of {\sc ROSAT} and my close friend in Germany.
}
\end{figure}

More detailed historical review was published in the year of Ya.~B.'s 100 anniversary
in ``Physics Uspekhi''~\cite{2014PhyU...57..407S}.

\section{Standard model of disk accretion}

Many excellent books and proceedings on accretion disk theory have been published so far.
I especially want to mention two of them.
Namely, ``Accretion Power in Astrophysics'' by Juhan Frank, Andrew King and
Derek Raine~\cite{2002apa..book.....F}
and ``Black Hole Accretion Disks. Towards a New Paradigm'' by Shoji Kato,
Jun Fukue and Sin Mineshige~\cite{2008bhad.book.....K}.
Bellow the theory of axisymmetric geometry thin and optically thick accretion disks
(a standard Shakura--Sunyaev $\alpha$-model) will be briefly outlined.

\subsection{Main equations}
The height-integrated continuity equation for the disk layer on some radius $r$:
\begin{equation}
 \frac{\partial\Sigma_0}{\partial t} = - \frac1{r} \frac{\partial}{\partial r}(\Sigma_0\,v_r\,r)\,,
 \label{cont}
\end{equation}
where $v_r$ is the radial velocity of the accreting matter,
$\Sigma_0 = \int_{-z_0}^{z_0}{\rho\,\mathrm{d}z}$ is surface density of accretion disk,
$\rho$ is the accreting matter density, $z_0$ is the semiheight of the disk and
$z$ and $t$ are coordinate along disk axis and time correspondingly.

The first Euler's equation:
\begin{equation}
 v_r \frac{\partial v_r}{\partial r} - \omega^2\,r = - \frac{G\,M}{r^2} - \frac1{\rho}\frac{\partial P}{\partial r} + ...
\end{equation}
Here $\omega$ is angular velocity of accreting mater,
$M$ is the mass of central object (i.e., black hole),
$P$ is the accreting matter pressure and ellipsis denotes other terms, such as viscosity,
small compared to main terms.
For geometrically thin accretion disks Keplerian rotation law takes place:
\begin{equation}
 \omega = \frac{v_\varphi}{r} = \sqrt{\frac{G\,M}{r^3}} \,.
\end{equation}

The second Euler's equation (the equation of angular momentum):
\begin{equation}
 \Sigma_0\,v_r \frac{\partial(\omega\,r^2)}{\partial r}
  = \frac1{r} \frac{\partial}{\partial r} (W_{r\varphi}\,r^2) \,,
 \label{vis}
\end{equation}
where $W_{r\varphi} = \int_{-z_0}^{z_0}w_{r\varphi}\,\mathrm{d}z$ is
height-integrated viscous shear stress.

\begin{figure}
\centering
\includegraphics[angle=0,width=0.45\textwidth]{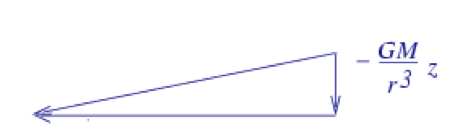}
\caption{
$z$-component of gravitational acceleration.
}
\label{fig.hydrostat}
\end{figure}

The equation of hydrostatic equilibrium along $z$-coordinate
(see Fig.~\ref{fig.hydrostat}):
\begin{equation}
 \frac1{\rho} \frac{\partial P}{\partial z} = -\frac{G\,M}{r^3} z \,.
 \label{dPdz}
\end{equation}

\subsection{The equation of the energy balance}
\begin{equation}
 \Sigma_0\,T\,v_r \frac{\partial S}{\partial r} = Q^+ - Q^- \,,
\end{equation}
where $T$ is the typical (along $z$-coordinate) accreting matter temperature,
$S$ is the entropy per unit mass,
$Q^+$ and $Q^-$ are the vertical-integrated rate of thermal energy release by
viscous shear and rate of thermal energy losses by radiation.
Left-hand side of the equation describes advection energy transfer,
for geometrically thin disks this value is small and $Q^+=Q^-$.

\subsection{Viscosity}
The main challenge of solving the disk accretion problem is to account for
viscous forces.
Usually, viscosity of ionized plasma is very low.
That is why it is necessary to assume the presence of developed turbulence
or magnetic fields in the accretion disk.
As the first step, let us introduce turbulence viscosity:
\begin{equation}
 w_{r\varphi}^t = \nu_t\,\rho\,r \frac{\partial \omega}{\partial r}
  =\rho <v_t\,l_t>r \frac{\partial \omega}{\partial r} \,,
\end{equation}
where $\nu_t$ is turbulent kinematic viscosity, $v_t$ and $l_t$ are turbulent speed and
length.

Using empirical Prandtl's law $v_t = l_t\,r |\partial\omega/\partial r|$ for Keplerian
disk we can rewrite:
\begin{equation}
 w_{r\varphi}^t = -\rho\,v_t^2 = - \alpha\,\rho\,v_s^2\,,
\end{equation}
where $v_s$ is the sound velocity of the accreting matter. Here we introduce the
$\alpha$-parameter:
\begin{equation}
 \alpha = (\mathcal{M}^t)^2 = \left( \frac{v_t}{v_s} \right)^2 \,,
\end{equation}
where $\mathcal{M}^t$ is the turbulent Mach number.

\subsection{Energy generation and transfer}

The energy release occurs due to the differential rotation of the viscous turbulent
disk, the energy release rate per unit volume is
\begin{equation}
 \epsilon = r\,w_{r\varphi} \frac{\partial \omega}{\partial r}
  = - \frac32 \,\omega\,w_{r\varphi} \,.
  \label{dqdz}
\end{equation}

The vertical gradient of the radiation flux $q$ is proportional to the energy
release rate:
\begin{equation}
 \frac{\partial q}{\partial z} = \epsilon \,.
 \label{dTdz}
\end{equation}

The equation transfer equation in the diffusion approximation:
\begin{equation}
 \frac{c}{3\,\varkappa\,\rho} \frac{\partial(a_r\,T^4)}{\partial z} = -q \,,
\end{equation}
where $\varkappa$ is the opacity coefficient and $a_r$ is the radiation constant.

\subsection{Vertical and radial disk structure}
Equations \eqref{dPdz},~\eqref{dqdz} and~\eqref{dTdz} with the continuity equation
in the form of $\partial \sigma / \partial z = \rho$ form the system of ordinary
differential equations, its solution for right boundary conditions provides
vertical disk structure.
This problem is like the problem of calculation of stellar structure, and can be solved
numerically, see, i.e.,~\cite{1998A&AT...15..193K,2017MNRAS.464..410M}.

In the case of stationary accretion disk,
when accretion rate $\dot{M} = 2\uppi\,\Sigma_0\,v_r\,r$ is constant,
we can integrate Eq.~\eqref{vis} and obtain radial dependence of viscous shear:
\begin{equation}
 W_{r\varphi}
  = \frac{\dot{M}\,\omega}{2\uppi} \left( 1 - \sqrt{\frac{R_\mathrm{in}}{r}} \right) \,,
\end{equation}
where $R_\mathrm{in}$ is the inner disk radius,
that equals innermost stable circular orbit for the case of accretion on the
black hole.
Thus, radial structure of accretion disk can be found as a solution of algebraic
equations~\cite{1973A&A....24..337S}.
Using the vertical structure solution analytical relations for radial distribution of various parameters of accretion disk can be obtained, see,
i.e.,~\cite{2000A&A...356..363L,2007ARep...51..549S}.

In the most simple case when radiation from the disk surface is black-body the
radiation flux is
\begin{equation}
 \begin{split}
 Q &= \frac12 r \frac{\partial \omega}{\partial r} W_{r\varphi}
  = \frac3{8\uppi} \dot{M} \frac{G\,M}{r^3} \left( 1 - \sqrt{\frac{R_\mathrm{in}}{r}} \right) \\
  &= \sigma_\mathrm{SB}\,T^4_\mathrm{eff} \,,
 \end{split}
\end{equation}
where $\sigma_\mathrm{SB}$ is the Stefan--Boltzmann constant.

The spectral radiative flux in a unit solid angle from a flat accretion disk at distance $d$ from the disk is equal to
\begin{equation}
 F_\nu
  = \frac{2\,\pi}{d^2} \, \cos {i}\, \int_{R_\mathrm{in}}^{R_\mathrm{out}} I_\nu \, r\,
\mathrm{d} r\, ,
\label{eq.disk_spectrum}
\end{equation}
where $\nu$ is radiation frequency, $i$ is the inclination of the disk to the line of sight, $R_\mathrm{out}$ is the outer disk radius and $I_\nu(r)$ is the intensity of radiation from the disk surface.

\begin{figure}
\centering
\includegraphics[angle=0,width=0.45\textwidth]{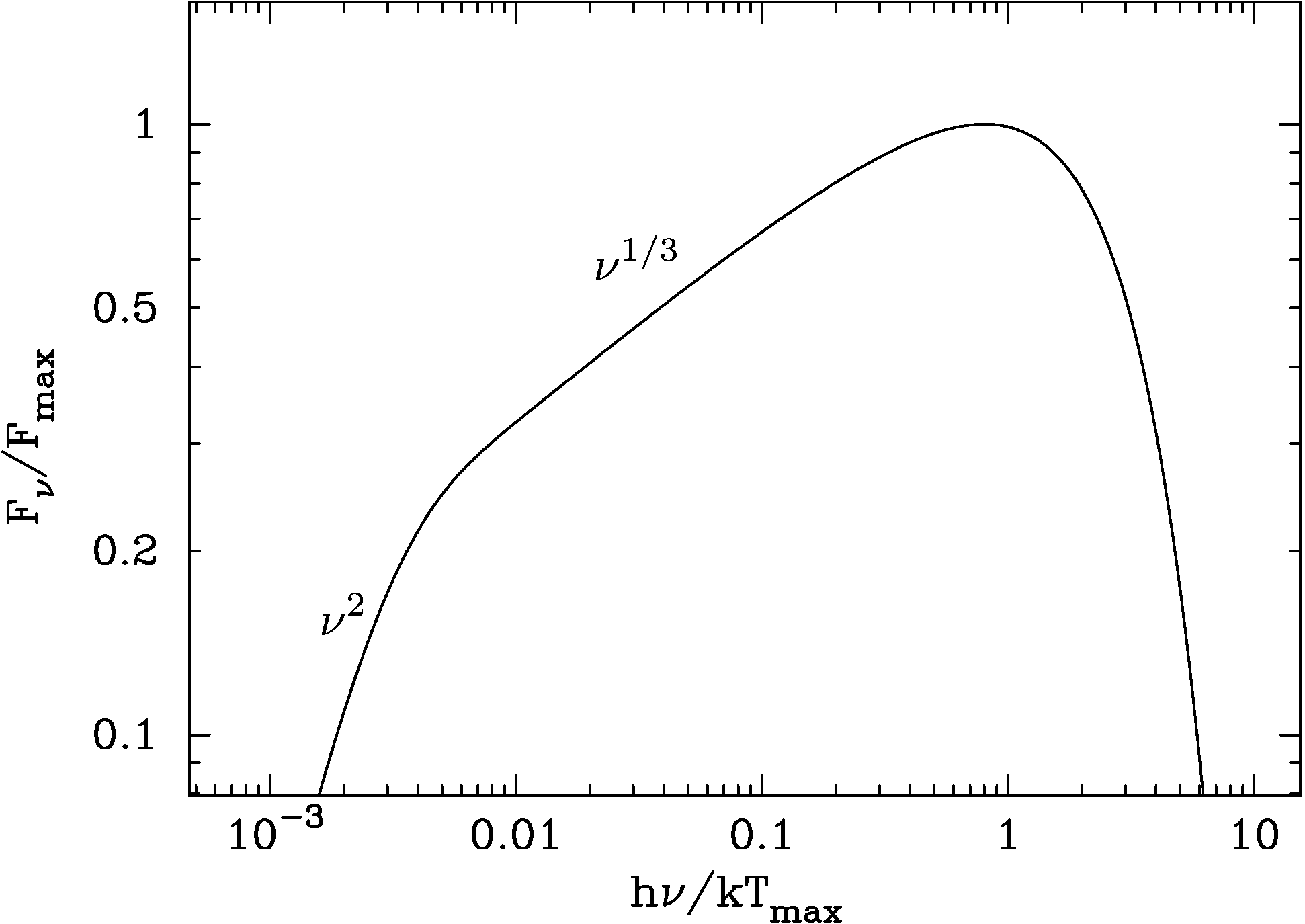}
\caption{
Spectral distribution of radiative flux density from a standard optically thick,
geometrically thin disk in the Newtonian metric.
The horizontal axis shows the normalised radiation frequency.
The vertical axis shows normalized spectral radiative flux density in units
of [erg/Hz/cm$^{2}$/s].
Here $h$ is the Planck constant, $k_\mathrm{B}$ is the Boltzmann constant, $T_\mathrm{max} = T_\mathrm{eff}\left(r = (7/6)^2\,R_\mathrm{in}\right)$ is the maximum surface temperature of the standard disk.
}
\label{fig.spectrum}
\end{figure}

With the assumption of black-body radiation Eq.~\eqref{eq.disk_spectrum} gives
spectral disk flux shown in Fig.~\ref{fig.spectrum}.
If the case of electron scattering spectrum has more complicated shape,
see~\cite{1986rpa..book.....R} for details.

\subsection{Non-stationary disk accretion}
Substituting Eq.~\eqref{vis} to Eq.~\eqref{cont} gives us equation of viscous disk
evolution. It is convenient to introduce a variable $F = -W_{r\varphi}$,
henceforth $2\,\uppi\,F$ means the total momentum of viscous forces acting
between the adjacent layers, and an independent variable
$h = \omega\,r^2$ (Keplerian angular momentum per unit mass).
Then we have:
\begin{equation}
 \frac{\dot{M}(r,t)}{2\,\uppi} = - \frac{\partial F}{\partial h}
 \label{dSdt}
\end{equation}
and
\begin{equation}
 \frac{\partial \Sigma_0}{\partial t}
  = \frac12 \frac{(G\,M)^2}{h^3} \frac{\partial^2 F}{\partial h^2} \,.
\end{equation}
A relation between $\Sigma_0$ and $F$ depends on the vertical structure, i.e. on opacity law, value of $\alpha$, etc. 
This equation is a non-linear partial differential equation of diffusion that in the general case can be
solved only numerically.

For the case of power-law opacity
$\varkappa(\rho, T) = \varkappa_0\,\rho^a/T^b$:
\begin{equation}
 \Sigma_0 = \frac{(G\,M)^2 F^{1-m}}{2(1-m)\,D\,h^{3-n}} \,,
\end{equation}
where $m$ and $n$ are dimensionless coefficients that depend on $a$ and $b$, and $D$
is dimension coefficient~\cite{2000A&A...356..363L}.
For this relation between $\Sigma_0$ and $F$ the equation of viscous
evolution~\eqref{dSdt} has the following form
\begin{equation}
 \frac{\partial F}{\partial t} = D \frac{F^m}{h^n} \frac{\partial^2 F}{\partial h^2}\,.
\end{equation}
There are a number of self-similar solution of this equation, see~\cite{1984AdSpR...3..305F,1987PAZh...13..917L,1988AdSpR...8..163F,2000A&A...356..363L}.
For the case of $m=0$ Green functions of this problem were found
in~\cite{2011MNRAS.410.1007T,2015ApJ...804...87L}.

\begin{figure}
\centering
\includegraphics[angle=0,width=0.45\textwidth]{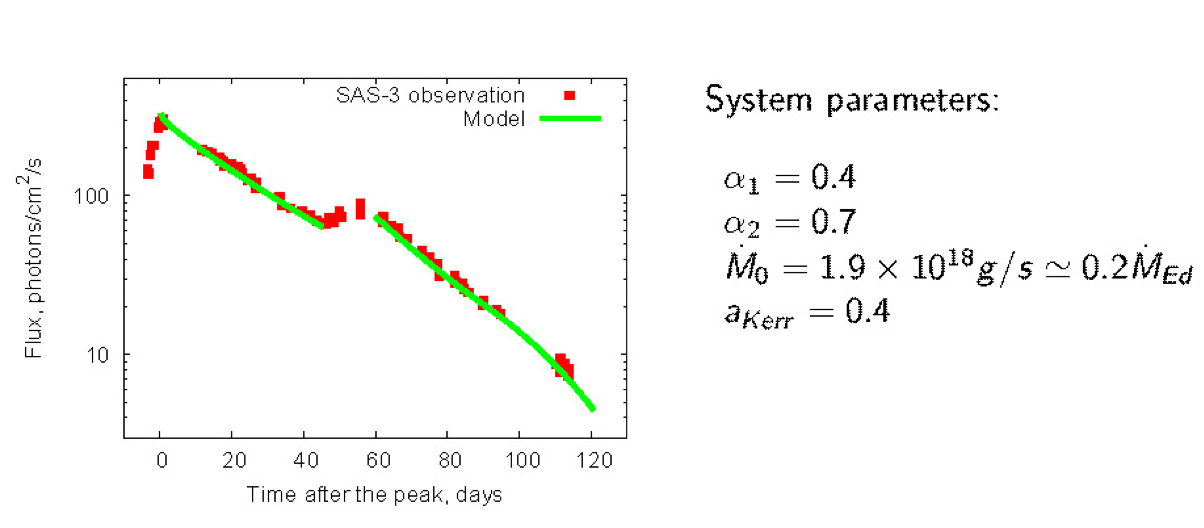}
\caption{
 X-ray light curve of 1975 outburst of A0620$-$00~\cite{2015AstL...41..797M}.
 Dots are observations, line is numerical model of accretion disk evolution.
}
\label{fig.0620}
\end{figure}

\begin{figure}
\centering
\includegraphics[angle=0,width=0.45\textwidth]{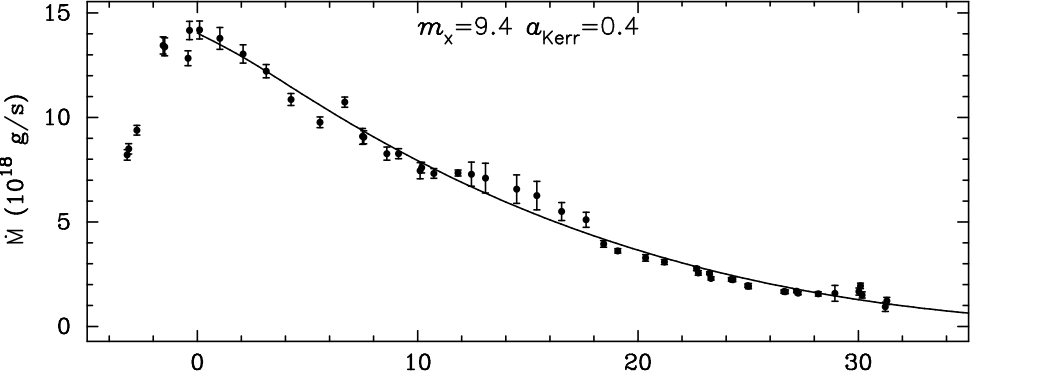}
\caption{
 Temporal evolution of accretion rate on the black hole in 2002 outburst of
 4U1543$-$47~\cite{2017MNRAS.468.4735L}.
}
\label{fig.4u1543}
\end{figure}

The theory of non-stationary accretion applies successfully to X-ray outbursts
of accretion disks in close binary systems with black holes or neutron stars.
Often such outbursts cannot be described in the terms of self-similar solutions of
the viscous equation~\eqref{dSdt}.
On the other hand numerical solution of this equation provides possibility to
take into account effects of disk self-irradiation and hydrogen recombination.
The examples of such numerical modeling are shown in Fig.~\ref{fig.0620} and~\ref{fig.4u1543}.
These simulations showed that $\alpha$-parameter is not small, it is about 0.7.

In the early 80-s Ya.~B. analysed the results of Sir Taylor's laboratory
experiment~\cite{1936RSPSA.157..546T} and explained generation of turbulence in
it~\cite{1981RSPSA.374..299Z}.
Several words about this experiment. Two coaxial cylinders were taken, the gap between
them was filled with some liquid.
The bigger cylinder was rotated with acceleration, the smaller one was stable.
At some moment the liquid became turbulent.
It was found that the bigger gap between cylinders required larger rotation velocity
to arise turbulence.
This relation was connected with critical Reynold's number.
Ya.~B.~Zeldovich suggested the theoretical basement which explained these
empirical data obtained in Taylor's experiment.
Nevertheless, Ya.~B. pointed out that this result cannot be applied to the disk
accretion.

\begin{figure}
\centering
\includegraphics[angle=0,width=0.45\textwidth]{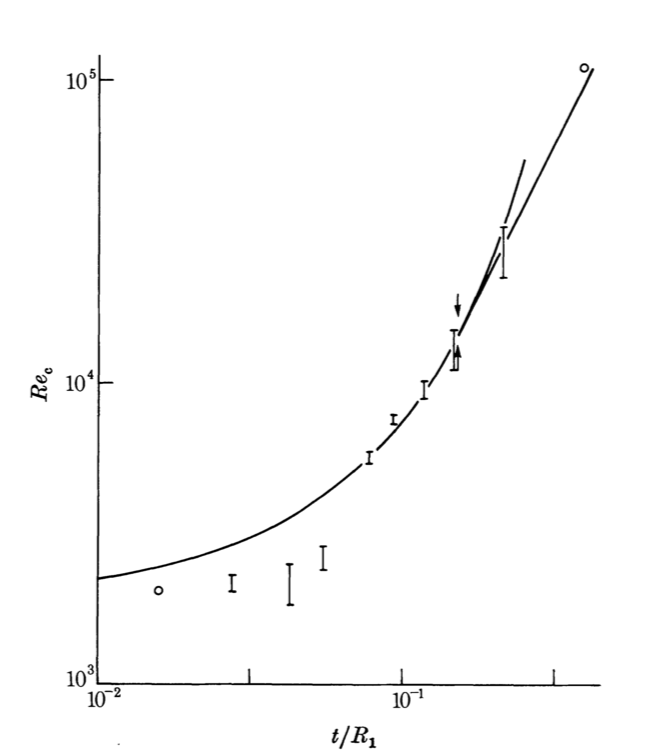}
\caption{
The critical Reynolds number dependence on the gap's width $t/R_1$.
Taylor's experimental results and the exponential approximation
by Ya.~B. are shown~\cite{1981RSPSA.374..299Z}.
}
\label{fig.taylor}
\end{figure}

The modern astrophysics is largely impacted by Zeldovich's papers on the interaction of
matter and radiation in the Universe and
on formation of large-scale structure of the Universe.
Yakov Borisovich Zeldovich died on the 2nd of December 1987.

I thank Konstantin Malanchev and Natalia Shakura for preparation of the manuscript.

\begin{minipage}{\linewidth}
\section{Instead Of Epilogue}
\begin{framed}
{
\centering \it
From the resolution of the Presidium of the Russian Academy of Sciences of 11 Feb. 2014
}

\begin{itemize}
 \item
 Establish the Academician Ya.~B.~Zeldovich gold medal, to be awarded by the Russian Academy of Sciences for distinguished works in physics and astrophysics.
 \item
Name a street in Moscow after Ya.~B.~Zeldovich.  
 \item
Set the commemorative plaques in memory of the Academician Ya.~B.~Zeldovich on the buildings of:
  \begin{itemize}
	\item the M.~V.~Keldysh Institute of Applied Mathematics of RAS
	\item the Space Research Institute of RAS
	\item the N.~N.~Semyonov Institute of Chemical Physics of RAS
  \end{itemize}
\end{itemize}
\end{framed}

\end{minipage}

{
\centering
\includegraphics[angle=0,width=0.45\textwidth]{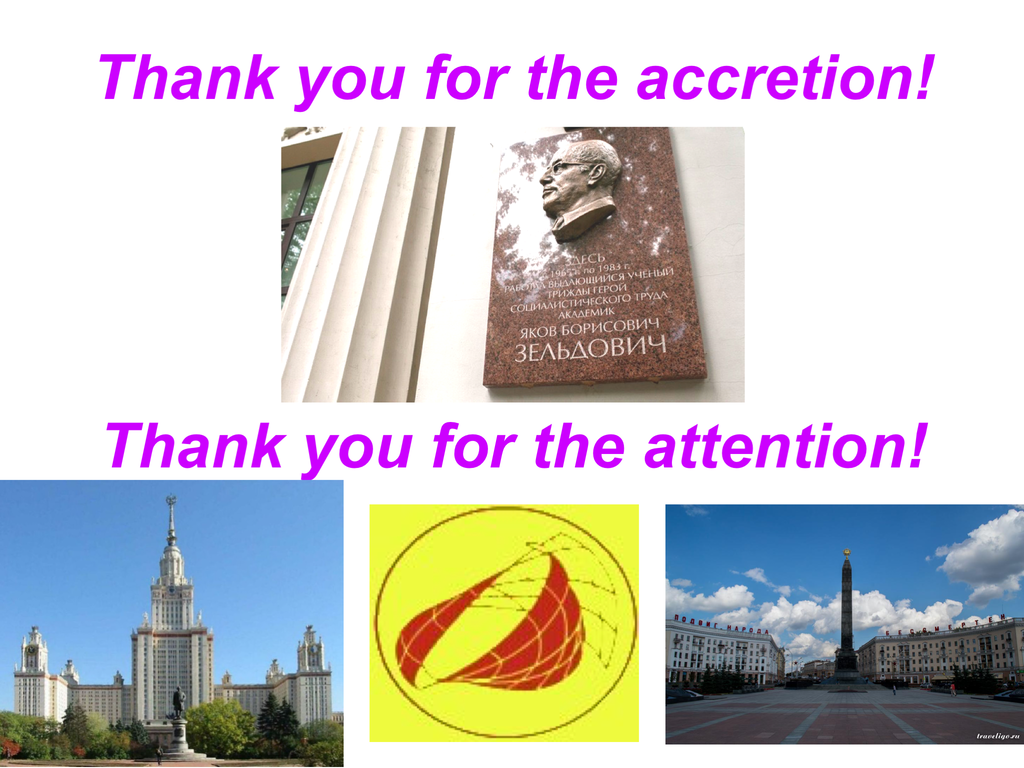}
}

\def\refitem#1{\relax}
\bibliographystyle{maik}
\bibliography{maikbibl}

\begin{thebibliography}{31}
\expandafter\ifx\csname natexlab\endcsname\relax\def\natexlab#1{#1}\fi
\expandafter\ifx\csname bibnamefont\endcsname\relax
  \def\bibnamefont#1{#1}\fi
\expandafter\ifx\csname bibfnamefont\endcsname\relax
  \def\bibfnamefont#1{#1}\fi
\expandafter\ifx\csname citenamefont\endcsname\relax
  \def\citenamefont#1{#1}\fi
\expandafter\ifx\csname url\endcsname\relax
  \def\url#1{\texttt{#1}}\fi
\expandafter\ifx\csname urlprefix\endcsname\relax\def\urlprefix{URL }\fi
\providecommand{\bibinfo}[2]{#2}
\providecommand{\eprint}[2][]{\url{#2}}

\bibitem[{\citenamefont{{Zeldovich} and
  {Khariton}}(1939)}]{1939ZhETF...9.1425Z}

\refitem{article}
\bibinfo{author}{\bibfnamefont{Y.~B.} \bibnamefont{{Zeldovich}}}
  \bibnamefont{and} \bibinfo{author}{\bibfnamefont{Y.~B.}
  \bibnamefont{{Khariton}}}, \bibinfo{journal}{Zhurnal Eksperimentalnoi i
  Teoreticheskoi Fiziki} \textbf{\bibinfo{volume}{9}}, \bibinfo{pages}{1425}
  (\bibinfo{year}{1939}).

\bibitem[{\citenamefont{{Zeldovich} and
  {Khariton}}(1940{\natexlab{a}})}]{1940ZhETF..10...29Z}

\refitem{article}
\bibinfo{author}{\bibfnamefont{Y.~B.} \bibnamefont{{Zeldovich}}}
  \bibnamefont{and} \bibinfo{author}{\bibfnamefont{Y.~B.}
  \bibnamefont{{Khariton}}}, \bibinfo{journal}{Zhurnal Eksperimentalnoi i
  Teoreticheskoi Fiziki} \textbf{\bibinfo{volume}{10}}, \bibinfo{pages}{29}
  (\bibinfo{year}{1940}{\natexlab{a}}).

\bibitem[{\citenamefont{{Zeldovich} and
  {Khariton}}(1940{\natexlab{b}})}]{1940ZhETF..10..477Z}

\refitem{article}
\bibinfo{author}{\bibfnamefont{Y.~B.} \bibnamefont{{Zeldovich}}}
  \bibnamefont{and} \bibinfo{author}{\bibfnamefont{Y.~B.}
  \bibnamefont{{Khariton}}}, \bibinfo{journal}{Zhurnal Eksperimentalnoi i
  Teoreticheskoi Fiziki} \textbf{\bibinfo{volume}{10}}, \bibinfo{pages}{477}
  (\bibinfo{year}{1940}{\natexlab{b}}).

\bibitem[{\citenamefont{{Hahn} and {Strassmann}}(1939)}]{1939NW.....27...11H}

\refitem{article}
\bibinfo{author}{\bibfnamefont{O.}~\bibnamefont{{Hahn}}} \bibnamefont{and}
  \bibinfo{author}{\bibfnamefont{F.}~\bibnamefont{{Strassmann}}},
  \bibinfo{journal}{Naturwissenschaften} \textbf{\bibinfo{volume}{27}},
  \bibinfo{pages}{11} (\bibinfo{year}{1939}).

\bibitem[{\citenamefont{{Meitner} and {Frisch}}(1939)}]{1939Natur.143..239M}

\refitem{article}
\bibinfo{author}{\bibfnamefont{L.}~\bibnamefont{{Meitner}}} \bibnamefont{and}
  \bibinfo{author}{\bibfnamefont{O.~R.} \bibnamefont{{Frisch}}},
  \bibinfo{journal}{\nat} \textbf{\bibinfo{volume}{143}}, \bibinfo{pages}{239}
  (\bibinfo{year}{1939}).

\bibitem[{\citenamefont{{Zeldovich} and {Shakura}}(1969)}]{1969AZh....46..225Z}

\refitem{article}
\bibinfo{author}{\bibfnamefont{Y.~B.} \bibnamefont{{Zeldovich}}}
  \bibnamefont{and} \bibinfo{author}{\bibfnamefont{N.~I.}
  \bibnamefont{{Shakura}}}, \bibinfo{journal}{\azh}
  \textbf{\bibinfo{volume}{46}}, \bibinfo{pages}{225} (\bibinfo{year}{1969}).

\bibitem[{\citenamefont{{Hewish} \emph{et~al.}}(1968)\citenamefont{{Hewish},
  {Bell}, {Pilkington}, {Scott}, and {Collins}}}]{1968Natur.217..709H}

\refitem{article}
\bibinfo{author}{\bibfnamefont{A.}~\bibnamefont{{Hewish}}},
  \bibinfo{author}{\bibfnamefont{S.~J.} \bibnamefont{{Bell}}},
  \bibinfo{author}{\bibfnamefont{J.~D.~H.} \bibnamefont{{Pilkington}}},
  \bibinfo{author}{\bibfnamefont{P.~F.} \bibnamefont{{Scott}}},
  \bibnamefont{and} \bibinfo{author}{\bibfnamefont{R.~A.}
  \bibnamefont{{Collins}}}, \bibinfo{journal}{\nat}
  \textbf{\bibinfo{volume}{217}}, \bibinfo{pages}{709} (\bibinfo{year}{1968}).

\bibitem[{\citenamefont{{Giacconi}}(1976)}]{1976ASIC...28..229G}

\refitem{inproceedings}
\bibinfo{author}{\bibfnamefont{R.}~\bibnamefont{{Giacconi}}}, in
  \emph{\bibinfo{booktitle}{NATO Advanced Science Institutes (ASI) Series C}},
  edited by \bibinfo{editor}{\bibfnamefont{G.}~\bibnamefont{{Setti}}}
  (\bibinfo{year}{1976}), vol.~\bibinfo{volume}{28} of
  \emph{\bibinfo{series}{NATO Advanced Science Institutes (ASI) Series C}}, pp.
  \bibinfo{pages}{229--269}.

\bibitem[{\citenamefont{{Shakura}}(1972)}]{1972AZh....49..921S}

\refitem{article}
\bibinfo{author}{\bibfnamefont{N.~I.} \bibnamefont{{Shakura}}},
  \bibinfo{journal}{\azh} \textbf{\bibinfo{volume}{49}}, \bibinfo{pages}{921}
  (\bibinfo{year}{1972}).

\bibitem[{\citenamefont{{Giacconi}
  \emph{et~al.}}(1971)\citenamefont{{Giacconi}, {Gursky}, {Kellogg},
  {Schreier}, and {Tananbaum}}}]{1971ApJ...167L..67G}

\refitem{article}
\bibinfo{author}{\bibfnamefont{R.}~\bibnamefont{{Giacconi}}},
  \bibinfo{author}{\bibfnamefont{H.}~\bibnamefont{{Gursky}}},
  \bibinfo{author}{\bibfnamefont{E.}~\bibnamefont{{Kellogg}}},
  \bibinfo{author}{\bibfnamefont{E.}~\bibnamefont{{Schreier}}},
  \bibnamefont{and}
  \bibinfo{author}{\bibfnamefont{H.}~\bibnamefont{{Tananbaum}}},
  \bibinfo{journal}{\apjl} \textbf{\bibinfo{volume}{167}}, \bibinfo{pages}{L67}
  (\bibinfo{year}{1971}).

\bibitem[{\citenamefont{{Shakura} and
  {Sunyaev}}(1973{\natexlab{a}})}]{1973IAUS...55..155S}

\refitem{inproceedings}
\bibinfo{author}{\bibfnamefont{N.~I.} \bibnamefont{{Shakura}}}
  \bibnamefont{and} \bibinfo{author}{\bibfnamefont{R.~A.}
  \bibnamefont{{Sunyaev}}}, in \emph{\bibinfo{booktitle}{X- and Gamma-Ray
  Astronomy}}, edited by
  \bibinfo{editor}{\bibfnamefont{H.}~\bibnamefont{{Bradt}}} \bibnamefont{and}
  \bibinfo{editor}{\bibfnamefont{R.}~\bibnamefont{{Giacconi}}}
  (\bibinfo{year}{1973}{\natexlab{a}}), vol.~\bibinfo{volume}{55} of
  \emph{\bibinfo{series}{IAU Symposium}}, p. \bibinfo{pages}{155}.

\bibitem[{\citenamefont{{Shakura} and
  {Sunyaev}}(1973{\natexlab{b}})}]{1973A&A....24..337S}

\refitem{article}
\bibinfo{author}{\bibfnamefont{N.~I.} \bibnamefont{{Shakura}}}
  \bibnamefont{and} \bibinfo{author}{\bibfnamefont{R.~A.}
  \bibnamefont{{Sunyaev}}}, \bibinfo{journal}{\aap}
  \textbf{\bibinfo{volume}{24}}, \bibinfo{pages}{337}
  (\bibinfo{year}{1973}{\natexlab{b}}).

\bibitem[{\citenamefont{{Novikov} and {Thorne}}(1973)}]{1973blho.conf..343N}

\refitem{inproceedings}
\bibinfo{author}{\bibfnamefont{I.~D.} \bibnamefont{{Novikov}}}
  \bibnamefont{and} \bibinfo{author}{\bibfnamefont{K.~S.}
  \bibnamefont{{Thorne}}}, in \emph{\bibinfo{booktitle}{Black Holes (Les Astres
  Occlus)}}, edited by
  \bibinfo{editor}{\bibfnamefont{C.}~\bibnamefont{{Dewitt}}} \bibnamefont{and}
  \bibinfo{editor}{\bibfnamefont{B.~S.} \bibnamefont{{Dewitt}}}
  (\bibinfo{year}{1973}), pp. \bibinfo{pages}{343--450}.

\bibitem[{\citenamefont{{Lynden-Bell}}(1969)}]{1969Natur.223..690L}

\refitem{article}
\bibinfo{author}{\bibfnamefont{D.}~\bibnamefont{{Lynden-Bell}}},
  \bibinfo{journal}{\nat} \textbf{\bibinfo{volume}{223}}, \bibinfo{pages}{690}
  (\bibinfo{year}{1969}).

\bibitem[{\citenamefont{{Shakura}}(2014)}]{2014PhyU...57..407S}

\refitem{article}
\bibinfo{author}{\bibfnamefont{N.~I.} \bibnamefont{{Shakura}}},
  \bibinfo{journal}{Physics Uspekhi} \textbf{\bibinfo{volume}{57}},
  \bibinfo{eid}{407-412} (\bibinfo{year}{2014}).

\bibitem[{\citenamefont{{Frank} \emph{et~al.}}(2002)\citenamefont{{Frank},
  {King}, and {Raine}}}]{2002apa..book.....F}

\refitem{book}
\bibinfo{author}{\bibfnamefont{J.}~\bibnamefont{{Frank}}},
  \bibinfo{author}{\bibfnamefont{A.}~\bibnamefont{{King}}}, \bibnamefont{and}
  \bibinfo{author}{\bibfnamefont{D.~J.} \bibnamefont{{Raine}}},
  \emph{\bibinfo{title}{{Accretion Power in Astrophysics: Third Edition}}}
  (\bibinfo{year}{2002}).

\bibitem[{\citenamefont{{Kato} \emph{et~al.}}(2008)\citenamefont{{Kato},
  {Fukue}, and {Mineshige}}}]{2008bhad.book.....K}

\refitem{book}
\bibinfo{author}{\bibfnamefont{S.}~\bibnamefont{{Kato}}},
  \bibinfo{author}{\bibfnamefont{J.}~\bibnamefont{{Fukue}}}, \bibnamefont{and}
  \bibinfo{author}{\bibfnamefont{S.}~\bibnamefont{{Mineshige}}},
  \emph{\bibinfo{title}{{Black-Hole Accretion Disks --- Towards a New Paradigm
  ---}}} (\bibinfo{year}{2008}).

\bibitem[{\citenamefont{{Ketsaris} and {Shakura}}(1998)}]{1998A&AT...15..193K}

\refitem{article}
\bibinfo{author}{\bibfnamefont{N.~A.} \bibnamefont{{Ketsaris}}}
  \bibnamefont{and} \bibinfo{author}{\bibfnamefont{N.~I.}
  \bibnamefont{{Shakura}}}, \bibinfo{journal}{Astronomical and Astrophysical
  Transactions} \textbf{\bibinfo{volume}{15}}, \bibinfo{pages}{193}
  (\bibinfo{year}{1998}).

\bibitem[{\citenamefont{{Malanchev}
  \emph{et~al.}}(2017)\citenamefont{{Malanchev}, {Postnov}, and
  {Shakura}}}]{2017MNRAS.464..410M}

\refitem{article}
\bibinfo{author}{\bibfnamefont{K.~L.} \bibnamefont{{Malanchev}}},
  \bibinfo{author}{\bibfnamefont{K.~A.} \bibnamefont{{Postnov}}},
  \bibnamefont{and} \bibinfo{author}{\bibfnamefont{N.~I.}
  \bibnamefont{{Shakura}}}, \bibinfo{journal}{\mnras}
  \textbf{\bibinfo{volume}{464}}, \bibinfo{pages}{410} (\bibinfo{year}{2017}),
  \eprint{1609.03799}.

\bibitem[{\citenamefont{{Lipunova} and {Shakura}}(2000)}]{2000A&A...356..363L}

\refitem{article}
\bibinfo{author}{\bibfnamefont{G.~V.} \bibnamefont{{Lipunova}}}
  \bibnamefont{and} \bibinfo{author}{\bibfnamefont{N.~I.}
  \bibnamefont{{Shakura}}}, \bibinfo{journal}{\aap}
  \textbf{\bibinfo{volume}{356}}, \bibinfo{pages}{363} (\bibinfo{year}{2000}),
  \eprint{astro-ph/0103274}.

\bibitem[{\citenamefont{{Suleimanov}
  \emph{et~al.}}(2007)\citenamefont{{Suleimanov}, {Lipunova}, and
  {Shakura}}}]{2007ARep...51..549S}

\refitem{article}
\bibinfo{author}{\bibfnamefont{V.~F.} \bibnamefont{{Suleimanov}}},
  \bibinfo{author}{\bibfnamefont{G.~V.} \bibnamefont{{Lipunova}}},
  \bibnamefont{and} \bibinfo{author}{\bibfnamefont{N.~I.}
  \bibnamefont{{Shakura}}}, \bibinfo{journal}{Astronomy Reports}
  \textbf{\bibinfo{volume}{51}}, \bibinfo{pages}{549} (\bibinfo{year}{2007}).

\bibitem[{\citenamefont{{Rybicki} and {Lightman}}(1986)}]{1986rpa..book.....R}

\refitem{book}
\bibinfo{author}{\bibfnamefont{G.~B.} \bibnamefont{{Rybicki}}}
  \bibnamefont{and} \bibinfo{author}{\bibfnamefont{A.~P.}
  \bibnamefont{{Lightman}}}, \emph{\bibinfo{title}{{Radiative Processes in
  Astrophysics}}} (\bibinfo{year}{1986}).

\bibitem[{\citenamefont{{Filipov}}(1984)}]{1984AdSpR...3..305F}

\refitem{article}
\bibinfo{author}{\bibfnamefont{L.~G.} \bibnamefont{{Filipov}}},
  \bibinfo{journal}{Advances in Space Research} \textbf{\bibinfo{volume}{3}},
  \bibinfo{pages}{305} (\bibinfo{year}{1984}).

\bibitem[{\citenamefont{{Lyubarskij} and
  {Shakura}}(1987)}]{1987PAZh...13..917L}

\refitem{article}
\bibinfo{author}{\bibfnamefont{Y.~E.} \bibnamefont{{Lyubarskij}}}
  \bibnamefont{and} \bibinfo{author}{\bibfnamefont{N.~I.}
  \bibnamefont{{Shakura}}}, \bibinfo{journal}{Pisma v Astronomicheskii Zhurnal}
  \textbf{\bibinfo{volume}{13}}, \bibinfo{pages}{917} (\bibinfo{year}{1987}).

\bibitem[{\citenamefont{{Filipov} \emph{et~al.}}(1988)\citenamefont{{Filipov},
  {Shakura}, and {Liubarskii}}}]{1988AdSpR...8..163F}

\refitem{article}
\bibinfo{author}{\bibfnamefont{L.}~\bibnamefont{{Filipov}}},
  \bibinfo{author}{\bibfnamefont{N.~I.} \bibnamefont{{Shakura}}},
  \bibnamefont{and}
  \bibinfo{author}{\bibfnamefont{I.}~\bibnamefont{{Liubarskii}}},
  \bibinfo{journal}{Advances in Space Research} \textbf{\bibinfo{volume}{8}},
  \bibinfo{pages}{163} (\bibinfo{year}{1988}).

\bibitem[{\citenamefont{{Tanaka}}(2011)}]{2011MNRAS.410.1007T}

\refitem{article}
\bibinfo{author}{\bibfnamefont{T.}~\bibnamefont{{Tanaka}}},
  \bibinfo{journal}{\mnras} \textbf{\bibinfo{volume}{410}},
  \bibinfo{pages}{1007} (\bibinfo{year}{2011}), \eprint{1007.4474}.

\bibitem[{\citenamefont{{Lipunova}}(2015)}]{2015ApJ...804...87L}

\refitem{article}
\bibinfo{author}{\bibfnamefont{G.~V.} \bibnamefont{{Lipunova}}},
  \bibinfo{journal}{\apj} \textbf{\bibinfo{volume}{804}}, \bibinfo{eid}{87}
  (\bibinfo{year}{2015}), \eprint{1503.09093}.

\bibitem[{\citenamefont{{Taylor}}(1936)}]{1936RSPSA.157..546T}

\refitem{article}
\bibinfo{author}{\bibfnamefont{G.~I.} \bibnamefont{{Taylor}}},
  \bibinfo{journal}{Proceedings of the Royal Society of London Series A}
  \textbf{\bibinfo{volume}{157}}, \bibinfo{pages}{546} (\bibinfo{year}{1936}).

\bibitem[{\citenamefont{{Zeldovich}}(1981)}]{1981RSPSA.374..299Z}

\refitem{article}
\bibinfo{author}{\bibfnamefont{Y.~B.} \bibnamefont{{Zeldovich}}},
  \bibinfo{journal}{Proceedings of the Royal Society of London Series A}
  \textbf{\bibinfo{volume}{374}}, \bibinfo{pages}{299} (\bibinfo{year}{1981}).

\bibitem[{\citenamefont{{Malanchev} and {Shakura}}(2015)}]{2015AstL...41..797M}

\refitem{article}
\bibinfo{author}{\bibfnamefont{K.~L.} \bibnamefont{{Malanchev}}}
  \bibnamefont{and} \bibinfo{author}{\bibfnamefont{N.~I.}
  \bibnamefont{{Shakura}}}, \bibinfo{journal}{Astronomy Letters}
  \textbf{\bibinfo{volume}{41}}, \bibinfo{pages}{797} (\bibinfo{year}{2015}),
  \eprint{1511.02356}.

\bibitem[{\citenamefont{{Lipunova} and
  {Malanchev}}(2017)}]{2017MNRAS.468.4735L}

\refitem{article}
\bibinfo{author}{\bibfnamefont{G.~V.} \bibnamefont{{Lipunova}}}
  \bibnamefont{and} \bibinfo{author}{\bibfnamefont{K.~L.}
  \bibnamefont{{Malanchev}}}, \bibinfo{journal}{\mnras}
  \textbf{\bibinfo{volume}{468}}, \bibinfo{pages}{4735} (\bibinfo{year}{2017}),
  \eprint{1610.01399}.

\end{thebibliography}

\end{document}